# Q-factor and emission pattern control of the WG modes in notched microdisk resonators

Svetlana V. Boriskina, *Member, IEEE*, Trevor M. Benson, *Senior Member, IEEE*,
Phillip Sewell, *Senior Member, IEEE*, and Alexander I. Nosich, *Fellow, IEEE*

*Abstract*— Two-dimensional (2-D) boundary integral equation analysis of a notched circular microdisk resonator is presented. Results obtained provide accurate description of optical modes, free from the staircasing and discretization errors of other numerical techniques. Splitting of the double degenerate Whispering-Gallery (WG) modes and directional light output is demonstrated. The effect of the notch depth and width on the resonance wavelengths, Q-factors, and emission patterns is studied. Further improvement of the directionality is demonstrated in an elliptical notched microdisk. Applications of the notched resonators to the design of microdisk lasers, oscillators, and biosensors are discussed.

## I. INTRODUCTION

Dielectric or semiconductor resonators shaped as circular cylinders and thin disks are, together with spherical particles, among the structures able to support the high-Q WG modes. Semiconductor microdisk lasers are very attractive light sources offering small mode volumes and ultralow threshold currents [1]. Perfectly circular microcavities provide very high optical confinement, which results in record Q-factors of the WG modes [2,3], however, they have two important drawbacks that limit their applications. These are, first, non-directive emission patterns with many identical beams, because a mode field in the disk plane depends on the azimuthal angle $\varphi$ as either $\cos m\varphi$ or $\sin m\varphi$ ($m$=0,1,2,…). Second, each mode with $m>0$ is double degenerate that leads to appearance of closely located doublets in the spectra of realistic resonators due to fabrication errors (sidewall roughness and shape imperfections, etc.) [3-5].

This work has been supported by the UK Engineering and Physical Sciences Research Council (EPSRC) under grants GR/R9055O/01 and GR/S60693/01P, and the Royal Society under grant IJP-2004/R1-FS.

The authors are with the George Green Institute for Electromagnetics Research, University of Nottingham, Nottingham NG7 2RD, UK (e-mail: eezsb@gwmail.nottingham.ac.uk or SBoriskina@gmail.com).

A. I. Nosich is also with the Institute of Radio Physics and Electronics NASU, Kharkov 61085, Ukraine.

To ensure a single-mode operation of the microdisk laser, it is desirable to stabilize the lasing mode against the fabrication imperfections [6] and either suppress all the parasitic modes (i.e., spoil their Q-factors) or detune their resonant frequencies away from that of the lasing mode [7]. As the lasing mode, we consider a fundamental transverse electric (TE) first-radial-order WG mode (one of the modes of a doublet) with the frequency at or near the spontaneous emission peak of the cavity material [2]. Several types of parasitic modes can be supported in a microdisk resonator, such as modes of the orthogonal (TM) polarization, higher-radial-order WG modes, and the other first-radial-order WG mode of a doublet.

TM-polarized emission is not usually observed in thin microdisks of several microns in diameter [3]. In high-index-contrast microdisks, the first radial-order WG-mode field is concentrated inside the microdisk very close to its rim. All of the higher-radial-order WG modes penetrate deeper inside the cavity. They can be suppressed by either decreasing the cavity radius and thus increasing their diffraction losses [8], or by removing material from the interior of the disk, which disturbs only the high-radial-order WG modes [7]. However, the former approach leads to increasing the diffraction losses of the lasing mode as well, and neither of them efficiently suppresses or shifts in frequency the second nearly degenerate first-radial-order WG mode of a doublet.

Recently, a suppression of such a parasitic mode using a circular microcavity with a rotationally periodic modification to its rim - a microgear laser cavity - has been reported [9]. Enhancement of the lasing WG mode Q factor in such a cavity enabled microgear lasers with low threshold currents to be fabricated [10]. However, for the microlaser applications, another important design parameter is the directionality of the light output [11]. The emission from a thin circular microdisk mostly occurs in the disk plane. Unfortunately, due to rotational symmetry of the circular microdisk or microgear resonators, lateral light directionality cannot be achieved. One of the ways to extract the light from the resonant cavity is to use output evanescent-field couplers of various geometries [12]. Alternatively, microcavity shape deformations that destroy the rotational symmetry can be introduced [13-16], which include elongation, projections, notches, and openings.

In this paper, we perform, for the first time to our knowledge, a detailed and accurate 2-D numerical study of the resonance and emission characteristics of a notched microdisk structure. We demonstrate efficient splitting and detuning of double-degenerate WG-modes as well as in-plane directional emission. As the microcavities of interest have wavelength-scale dimensions and regions of high contour curvature, conventional optical ray-tracing methods such as paraxial approximation and billiard theory fail to provide reliable results. Therefore, in this paper we analyze a microdisk resonator with a narrow notch using a 2-D Muller boundary integral equation (MBIE) formulation and the trigonometric-Galerkin discretization method [17]. Unlike the FDTD techniques, which imply extraction of the resonance frequencies from the transient field due to pulsed source excitation [3,9], this accurate, reliable and versatile method enables a direct a study of the resonance spectra, Q-factors, and emission patterns of arbitrarily shaped 2-D microcavities.

## II. NOTCHED MICRODISK GEOMETRY

Fig. 1 presents the in-plane geometry of a semiconductor microdisk with a narrow notch and the coordinate system used in the analysis. We consider a 2-D model of the structure in the *x-y* plane, accounting for the microdisk thickness by using the Effective Index Method. Here, the effective refractive index of the 2-D microcavity is taken as the normalized propagation constant of the fundamental guided mode in an equivalent planar waveguide at a wavelength corresponding to the spontaneous emission peak of the cavity material at room temperature [17].

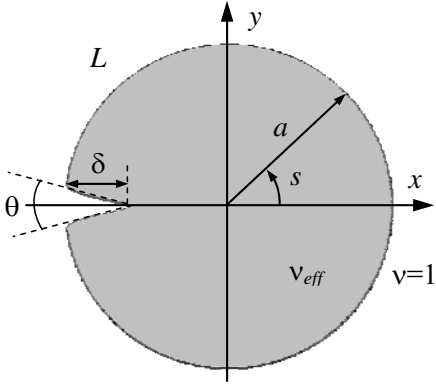

Fig. 1. Schematic of a 2-D geometry of a circular microcavity with a notch. The notch causes splitting of the double-degenerate WG modes.

In the 2-D model, the microcavity can support the modes of two polarizations: transverse electric (TE) with the electric field vector parallel to the *xy* plane and transverse magnetic (TM) with the same vector perpendicular to the plane of the microdisk. In a thin microdisk whose thickness is a small fraction of the optical wavelength, quasi-TE-polarized modes are dominant due to much larger effective index values [2]. Therefore, in our numerical analysis we will consider only these modes. We will also use the value of the effective refractive index of a slightly lossy microcavity given by $\nu_{eff} = 2.63 + i10^{-4}$. The real part of this value corresponds to the normalized propagation constant of the fundamental mode in a 200 nm-thick slab of GaInAsP (bulk refractive index 3.37) at 1.55 $\mu$m [3,10].

The contour of the microdisk cross-section by the plane (*x,y*) is characterized by a smooth 2-D closed curve *L*, which can be presented in the parametrical form as follows:

$$x = ar(s)\cos s, \quad y = ar(s)\sin s, \quad 0 \le s \le 2\pi, \quad (1)$$

where

$$r(s) = 1 - \delta/2 \cdot (\cos(2\tau s) + 1), \quad \pi(1 - 1/2\tau) < s < \pi(1 + 1/2\tau)$$
$$r(s) = 1, \quad 0 \le s \le \pi(1 - 1/2\tau) \text{ and } \pi(1 + 1/2\tau) \le s \le 2\pi \quad (2)$$

Here, *a* is a microdisk radius, $\delta$ is a notch depth, $\theta = \pi/\tau$ is a notch angular width, and parameter *s* is the polar angle.

## III. PROBLEM FORMULATION AND SOLUTION

In the 2-D formulation, the total field can be characterized by a single scalar function, which represents either $E_z$ or $H_z$ component in the case of the TM or TE polarization, respectively. This function satisfies the Helmholtz equation together with continuity conditions on contour *L*, and can be reduced to the following set of the second-kind boundary IEs with integrable kernels [17,18]:

$$U(\vec{r}) = \int_L \left[ U(\vec{r}\,') \left( \frac{\partial G_d}{\partial n'} - \frac{\partial G}{\partial n'} \right) - V(\vec{r}\,') \left( G_d - \frac{1}{\alpha_d} G \right) \right] dl' \quad (3)$$

$$\frac{1+\alpha}{2\alpha} V(\vec{r}) =$$
$$\int_L \left[ U(\vec{r}\,') \left( \frac{\partial^2 G_d}{\partial n \partial n'} - \frac{\partial^2 G}{\partial n \partial n'} \right) - V(\vec{r}\,') \left( \frac{\partial G_d}{\partial n} - \frac{1}{\alpha_d} \frac{\partial G}{\partial n} \right) \right] dl' \quad (4)$$

Here, the unknowns *U* and *V* are the limiting values of the field function and its normal derivative, respectively, from inside of the contour *L*, $\partial/\partial n$ is the normal derivative, and $\vec{n}$ and $\vec{n}'$ are the inward normal unit vectors to *L* at the observation and source points, respectively. The coefficient $\alpha$ is equal to either 1 in the TM-polarization case or $\nu_{eff}^2$ in the TE-polarization case. Time dependence is adopted as $\exp(-i\omega t)$ and is omitted throughout the paper. Functions *G* and $G_d$ in the kernels of IE (4) are given by

$$G_d(\vec{r}, \vec{r}\,') = \frac{i}{4} H_0^{(1)}(k\nu_{eff}|\vec{r} - \vec{r}\,'|),$$
$$G(\vec{r}, \vec{r}\,') = \frac{i}{4} H_0^{(1)}(k|\vec{r} - \vec{r}\,'|) \quad (5)$$

Applying the trigonometric Galerkin discretization method together with extraction and analytical integration of the kernel singularities (see [17] for details), the IEs (3,4) are converted into the following homogeneous matrix equation:

$$a_m^{11}U_m + a_m^{12}V_m + \sum_{(n)}\left(U_n A_{mn}^{11} + V_n A_{mn}^{12}\right) = 0$$
$$a_m^{21}U_m + a_m^{22}V_m + \sum_{(n)}\left(U_n A_{mn}^{21} + V_n A_{mn}^{22}\right) = 0 \quad , \qquad (6)$$

where

$$a_m^{11} = \nu_{eff} J_m^d H_m^{d\prime} - J_m' H_m + 2/i\pi\kappa$$
$$a_m^{12} = J_m^d H_m^d - J_m H_m/\alpha$$
$$a_m^{21} = J_m' H_m' - \nu_{eff}^2 J_m^{d\prime} H_m^{d\prime} \qquad (7)$$
$$a_m^{22} = J_m H_m'/\alpha - \nu_{eff} J_m^{d\prime} H_m^d + (1+\alpha)/i\pi\kappa\alpha$$

Here we denoted $\kappa = ka$, $J_m = J_m(\kappa)$, $J_m^d = J_m(\kappa\nu_{eff})$, $H_m = H_m^{(1)}(\kappa)$ and $H_m^d = H_m^{(1)}(\kappa\nu_{eff})$ are the Bessel and Hankel functions, respectively, and the prime represents the differentiation in argument. The matrix coefficients are double Fourier-type integrals of regular functions evaluated numerically via the Fast Fourier Transform algorithm [17]:

$$A_{mn}^{11} = 2L_{m-n}/i\pi - \int_0^{2\pi}\int_0^{2\pi}\left(\frac{\partial G_d}{\partial n'} - \frac{\partial G}{\partial n'} - \frac{\partial G_d^c}{\partial n'} + \frac{\partial G^c}{\partial n'}\right) e^{ins' - ims} ds\,ds' \qquad (8)$$

$$A_{mn}^{12} = \int_0^{2\pi}\int_0^{2\pi}\left(G_d - \frac{1}{\alpha_d}\cdot G - G_d^c + \frac{1}{\alpha_d}\cdot G^c\right) e^{ins' - ims} ds\,ds' \qquad (9)$$

$$A_{mn}^{21} = -\int_0^{2\pi}\int_0^{2\pi}\left(\frac{\partial^2 G_d}{\partial n\partial n'} - \frac{\partial^2 G}{\partial n\partial n'} - \frac{\partial^2 G_d^c}{\partial n\partial n'} + \frac{\partial^2 G^c}{\partial n\partial n'}\right) e^{ins' - ims} ds\,ds' \qquad (10)$$

$$A_{mn}^{22} = (1+\alpha_d)L_{m-n}/i\pi\alpha_d + \int_0^{2\pi}\int_0^{2\pi}\left(\frac{\partial G_d}{\partial n} - \frac{1}{\alpha_d}\cdot\frac{\partial G}{\partial n} - \frac{\partial G_d^c}{\partial n} + \frac{1}{\alpha_d}\cdot\frac{\partial G^c}{\partial n}\right) e^{ins' - ims} ds\,ds' \qquad (11)$$

where

$$L_m = \frac{1}{2\pi}\int_0^{2\pi}\left((dx/ds)^2 + (dy/ds)^2\right)^{-1/2} e^{-ims} ds. \qquad (12)$$

and $G^c$ and $G_d^c$ are the values of the functions (5) on the circle of radius $a$.

Finally, by introducing new unknowns: $z_m^1 = a_m^{11}U_m + a_m^{12}V_m$, $z_m^2 = a_m^{21}U_m + a_m^{22}V_m$, (6) can be reduced to the following final canonical form 2×2 block-type infinite-matrix equation of the Fredholm second kind:

$$\begin{bmatrix}\mathbf{z}^1\\\mathbf{z}^2\end{bmatrix} + \begin{bmatrix}\mathbf{M}^{11} & \mathbf{M}^{12}\\\mathbf{M}^{21} & \mathbf{M}^{22}\end{bmatrix}\cdot\begin{bmatrix}\mathbf{z}^1\\\mathbf{z}^2\end{bmatrix} = 0. \qquad (13)$$

The homogeneous matrix equation (13) has nontrivial solutions only at discrete complex values of the dimensionless frequency parameter $\kappa$, where the determinant of the matrix is zero. As we assume the time dependence $\exp(-i\omega t)$, $\mathrm{Im}\,\kappa$ can only have negative values. The search for the complex roots of the determinant equation was performed by means of the Powell hybrid method, and yielded both the resonance wavelengths and the quality factors of the microdisk modes:

$$\lambda = 2\pi a/\mathrm{Re}\,\kappa\,, \quad Q = -\mathrm{Re}\,\kappa/2\,\mathrm{Im}\,\kappa\,. \qquad (14)$$

After a complex natural frequency is found, the near and far-field patterns of the corresponding mode can be calculated within a multiplicative constant, through the corresponding solution to (13).

## IV. MODE SPLITTING IN THE NOTCHED MICRODISK

TE-polarized modes of ideal circular microdisk are usually classified as $E_{mnq}$ modes, where the subscripts $m$, $n$ and $q$ correspond to the number of azimuthal, radial, and off-plane variations of the mode field, respectively. A mode displays a "whispering-gallery" behavior due to nearly total internal reflection and $\mathrm{Im}\,\kappa \approx -const\,e^{-\mathrm{Re}\,\kappa}$ only if $k/\nu_{eff} < m < \kappa$ (see [19,20] for details). Besides, we imply here that for all the modes supported by thin microdisks $q=0$ and thus omit this index. All the WGE$_{mn}$ modes in circular microdisks with $m > 0$ are double degenerate (with either $\cos(ms)$ or $\sin(ms)$ angular field dependence) due to the microdisk rotational symmetry. We shall denote these two orthogonal states of the same mode as S-mode and A-mode, respectively, having either symmetrical or anti-symmetrical field patterns with respect to $s=0$.

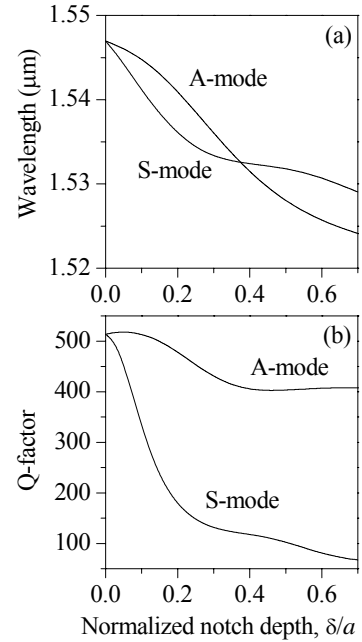

Fig. 2. (a) Resonance wavelengths and (b) Q-factors of the symmetrical (S-) and asymmetrical (A-) WGE$_{6,1}$ modes in a 1.8-μm diameter notched microdisk as a function of the notch depth.

Note that in the case of the notch shape given by (2), the contour $L$ has a line of symmetry (the $x$-axis), therefore it is convenient to count the angle *s* from this line. When the contour is deformed from a circle, the S-mode and the A-mode experience different shifts in their complex natural frequencies, and the degeneracy is removed. For clarity, we shall classify these two non-degenerate modes of the perturbed structure with the same indices *m* and *n* as in the unperturbed case.

In the vicinity of the spontaneous emission peak in a 1.8 $\mu$m-diameter GaInAsP circular microdisk, we find a double-degenerate $WGE_{6,1}$ mode with $\lambda = 1.547$ $\mu$m and Q = 513. In Fig. 2, we plot the mode wavelengths and Q-factors as a function of the notch depth. Fig. 2a demonstrates that making a notch causes splitting of the double-degenerate $WGE_{6,1}$ mode into two modes of orthogonal symmetry.

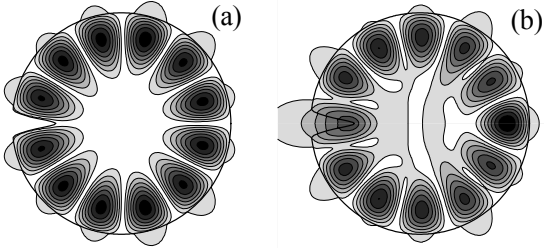

Fig. 3. Near-field portraits (12.5% contours) of (a) A-mode ($\lambda = 1.531$ $\mu$m) and (b) S- mode ($\lambda = 1.532$ $\mu$m) of the notched microdisk with the same parameters as in Fig. 2 and the normalized notch depth $\delta/a = 0.4$.

In a previous publication [5], we showed that efficient manipulation of the mode wavelength and Q-factor could be achieved if a contour deformation periodicity is matched to the modal field pattern. Although the cavity shown in Fig. 1 has a localized rather then periodical deformation, similar matching can be foreseen. Therefore, the notch width was chosen as a half of the distance between a neighboring maximum and minimum in the unperturbed $WGE_{6,1}$ mode field pattern ($\tau = 12$). Such a localized contour deformation is expected to have more significant effect on the S-mode, which has a field maximum on the *x*-axis, i.e., in the region of the notch, than on the A-mode, which has a zero field at the same location. Indeed, it can be seen in Fig. 2b that the Q-factor of the S-mode is noticeably decreased in the notched microdisk, while that of the A-mode remains almost as high as in the ideal circular resonator.

The field portraits (equal value curves of $|H_z(x,y)|$) of the A-mode and S-mode in the notched microdisk are presented in Fig. 3. As expected, the A-mode near-field pattern is not visibly affected by the notch, while the S-mode field pattern is clearly distorted. Because of the much higher energy leakage at the region of the notch, the S-mode Q-factor is lower that that of the A-mode.

Efficient separation of two resonant wavelengths in the notched microdisk, together with the spoiling of the Q-factor of one mode of a doublet, is crucially important in many applications of resonators with WG modes. Due to fabrication imperfections, circular disk resonators spectra often display two closely spaced resonances around each WM mode wavelength. This causes mode hopping and parasitic losses. A modified resonator design that avoids two equally coupled closely located WG modes may help achieve a quasi-single-mode operation of microdisk lasers, optical waveguide filters [21] and oscillators for satellite communications [22].

## V. STABILITY OF THE HIGH-Q A-MODE

We shall now study how the variations in the notch width and depth may affect the resonance wavelength and quality factor of the high-Q anti-symmetrical mode. Fig. 4 shows the A-mode wavelength detuning (a) and Q-factor change (b) with an increase of the notch depth for three values of the notch width. As expected, the wider the notch the more noticeable mode wavelength detuning and Q-factor decrease. However, a general behavior of the graphs in Fig. 4 is the same for all values of the notch width, and even in the case of the widest notch ($\tau = 10$) the A-mode Q-factor remains relatively high (~70% of that of the circular microdisk).

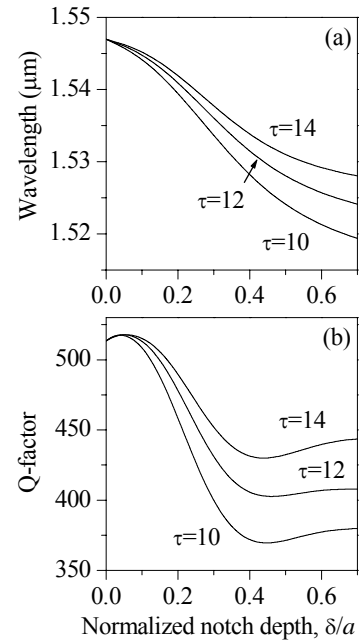

Fig. 4. (a) Resonance wavelengths and (b) quality factors of the anti-symmetrical $WGE_{6,1}$ mode in the 1.8-$\mu$m diameter notched microdisk as a function of the notch depth for three different values of the notch width.

Naturally, to efficiently split higher-azimuthal-order WG modes without spoiling the A-mode Q-factor, narrower notches will be required (the larger the azimuthal mode number the narrower the notch). It should be noted here that introducing notches whose width is not matched to the WG-mode field patterns (too wide) affects the modes of both symmetries almost equally (similarly to periodic contour corrugations or symmetrical shape deformations [4,5]). Namely, it may significantly blueshift both modes and spoil their Q-factors, instead of efficiently tuning their wavelengths away from each other.

Another important design parameter that depends on the precision of a fabrication procedure is the notch depth. The variations in the notch depth can result in unpredictable detuning of the wavelengths of both modes. However, it can be seen in Fig. 4a that with the increase of the notch depth the values of resonance wavelengths change more rapidly for shallow than for deep notches. Moreover, after first decreasing significantly, the values of Q-factors slightly increase and then flatten with further increase of the notch depth (Fig. 4b).

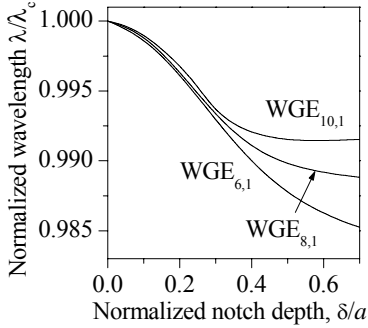

Fig. 5. Resonance wavelengths of the anti-symmetrical $WGE_{6,1}$, $WGE_{8,1}$, $WGE_{10,1}$ modes in notched microdisks with diameters 1.8 μm, 2.2 μm and 2.7 μm, respectively, as a function of the notch depth. The wavelengths are normalized to the corresponding wavelengths of the circular microdisks: $\lambda_c(WGE_{6,1})$ = 1.547 μm, $\lambda_c(WGE_{8,1})$ = 1.521 μm, $\lambda_c(WGE_{10,1})$ = 1.569 μm.

By looking at the $WGE_{6,1}$ mode field portraits (Fig. 3), one can notice that the mode field penetrates rather far towards the center of the microdisk. It is well known, however, that the electromagnetic fields of WG modes with $m \gg 1$ are stronger confined to the rim of the resonator. Therefore one can expect similar graphs of their resonance wavelengths to flatten at smaller values of the notch depth.

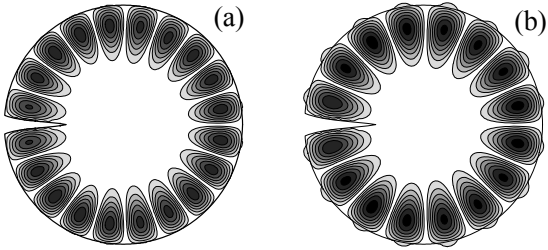

Fig. 6. Near-field portraits (12.5% contours) of anti-symmetrical (a) $WGE_{10,1}$ ($\lambda$ = 1.555 μm) and (b) $WGE_{8,1}$ ($\lambda$ = 1.505 μm) modes in notched microdisks with the same parameters as in Fig. 5, and the normalized notch depths $\delta/a$ = 0.5 and $\delta/a$ = 0.6, respectively.

In Fig. 5, the resonance wavelength detuning with the increase of the notch depth is plotted for three A-WG modes: $WGE_{6,1}$, $WGE_{8,1}$, and $WGE_{10,1}$. Though for all the three modes the wavelength values are first changing rapidly, then they begin to stabilize at a certain level. Notice that this level is reached faster for the modes with larger $m$. The near fields for the A-$WGE_{10,1}$ ($\delta/a$ = 0.5) and A-$WGE_{8,1}$ ($\delta/a$ = 0.6) modes are plotted in Fig. 6. The conclusion is that if the notch is deep enough to pierce through the area of the WG-mode field concentration, slight variations in its depth do not cause noticeable detuning of the A-mode.

## VI. EMISSION DIRECTIONALITY

Having found the complex natural frequencies of (13) we can compute the far-field emission patterns as well as the near-field portraits. In the far zone of the microdisk ($r \to \infty$), the field function can be presented in the following form [17]:

$$U(\vec{r}) = (1/kr)^{1/2} \exp(ikr)\Phi(\varphi), \qquad (15)$$

where $\Phi(\varphi)$ is the far-field emission pattern and $\varphi$ is the observation angle.

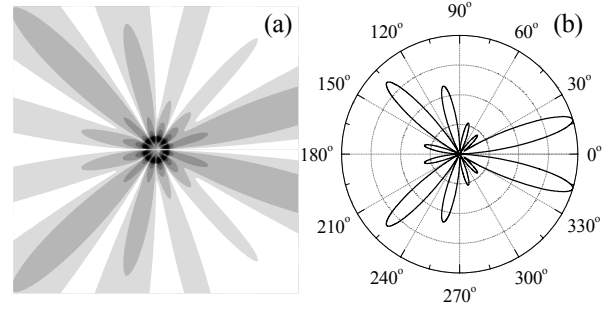

Fig. 7. (a) Near-field portrait and (b) far-field emission pattern of the anti-symmetrical $WGE_{6,1}$ mode ($\lambda$ = 1.531 μm) in the 1.8-μm diameter notched microdisk with the normalized notch depth $\delta/a$ = 0.4.

First, we calculated the near and the far-field patterns of the high-Q anti-symmetrical $WGE_{6,1}$ mode (Fig. 7). The A-mode emission pattern is found to consist of twelve beams, similarly to its counterpart for a perfectly circular cavity. However, the beams are no more identical, i.e., emission into some of the beams is more intense than into others. Still it is clear that the notched microdisk laser source operating on the A-mode will not emit light unidirectionally – at least two equal main beams are always present. Better control of the emission directionality can be achieved with the S-mode, whose modal pattern is not zero along the symmetry axis both in perfect and notched disks (Fig. 3b).

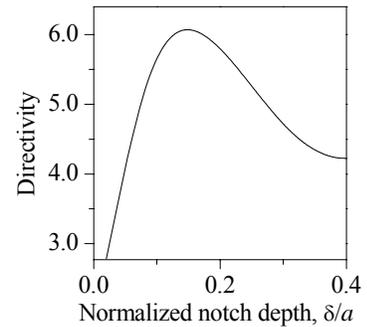

Fig. 8. Directivity of the symmetrical $WGE_{6,1}$ mode emission in the 1.8-μm diameter notched microdisk as a function of the notch depth.

The degree of collimation of the emitted light can be measured in terms of directivity. This quantity is well known in antenna theory and is defined as the ratio of the intensity of light radiated in the main-beam direction $\varphi_0$ to the intensity averaged over all directions:

$$D = 2\pi |\Phi(\varphi_0)|^2 \cdot \left( \int_0^{2\pi} |\Phi(\varphi)|^2 d\varphi \right)^{-1} \quad (16)$$

The higher the value of directivity, the better the light is collimated into a single directional beam in the far zone of the microdisk.

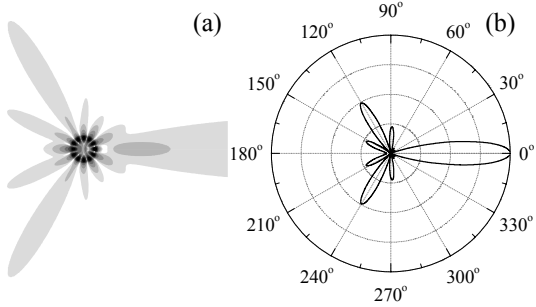

Fig. 9. (a) Near-field portrait and (b) far-field emission pattern of the symmetrical WGE$_{6,1}$ mode ($\lambda$ = 1.539 μm) in the notched microdisk with the normalized notch depth $\delta/a$ = 0.14, corresponding to the maximum of directivity in Fig. 8.

We expect to control the WG-mode emission pattern and enhance the directivity by tuning the notch geometry, with the aim to design light sources with narrow directional emission patterns. Fig. 8 shows how the value of directivity of the symmetrical S-WGE$_{6,1}$ mode varies with the increase of the notch depth. The emission pattern has the main beam at $\varphi=0^o$ (Fig. 9b) and a number of sidelobes. (Note, that the highest intensity in the near-field distribution of the S-mode is observed at the region of the notch, i.e., at $\varphi=180^o$). The directivity of emission increases rapidly with increasing the notch depth, reaching a maximum at $\delta/a$=0.14. Near-field portrait and far-field emission pattern shown in Fig. 9 are calculated for the notch depth corresponding to the maximum of directivity in Fig. 8. Note, however, that the S-mode Q-factor is five times lower than the Q-factor of the A-mode (see Fig. 2b), which normally leads to a higher threshold of lasing [19,20].

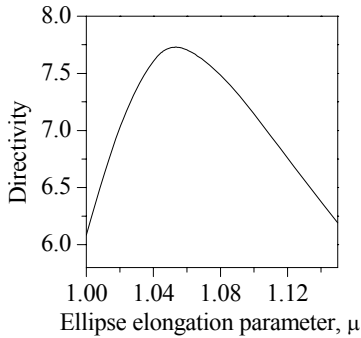

Fig. 10. Directivity of the symmetrical WGE$_{6,1}$ mode emission in an *elliptical* notched microdisk with the minor axis length 0.9 μm, and the normalized notch depth $\delta/a$ = 0.14 as a function of the ellipse elongation parameter μ.

Though the S-mode emission is directional, it is desirable to further reduce the sidelobes level. It has been observed theoretically and experimentally [4,8,13], that emission from the elliptical microdisks supporting distorted WG-modes is more directional than from circular ones. In elliptical resonators, WG-mode emission occurs at the points of the highest curvature of the contour and collimates into a number of beams with the highest-intensity beams forming around the ellipse minor axis (i.e., at $\varphi=0^o$ and $180^o$ for the ellipse elongated along the y-axis). Such improvement of emission directionality with increasing the ellipticity of the microdisk up to a certain critical value has been demonstrated experimentally in Ref. 8.

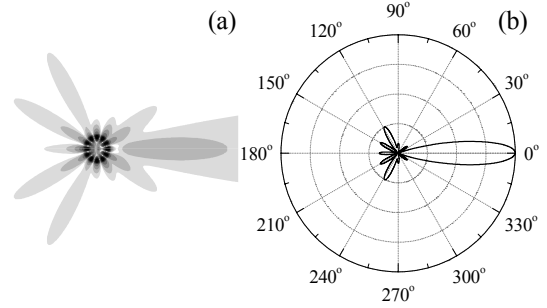

Fig. 11. (a) Near-field portrait and (b) far-field emission pattern of the symmetrical WGE$_{6,1}$ mode ($\lambda$ = 1.578 μm) in the elliptical notched microdisk with the normalized notch depth $\delta/a$ = 0.14 and the ellipse elongation parameter μ = 1.05, corresponding to the maximum of directivity in Fig. 10.

With this in mind, we expect to further enhance the directivity of the emission pattern by adjusting both the notch depth and the ellipticity of the resonator. Simulation results obtained with the same algorithm are presented in Figs. 10. By varying the elongation parameter of the microdisk, μ (the ratio of the major to the minor axis length), we observe a maximum of directivity at μ=1.05. The near and far field patterns plotted in Fig. 11 for such a cavity clearly show a highly directional emission with weaker sidelobes.

## VII. CONCLUSIONS

Results from the Muller boundary integral equations analysis of notched microcavities have been presented that provide clear insight into their improved optical performance over its smooth microdisk equivalent. Proposed notched resonator design provides efficient control of both frequency separation and Q-factors of two symmetry types of originally double-degenerate WG-modes, as well as directional light output. The directivity of emission can be further improved by distorting the microdisk shape from circular to elliptical one, although in general the demands of the high Q-factor and high directivity are contradictory. Applications of the notched disk resonators supporting non-degenerate first radial-order high-Q WG-modes are evident. They are expected to have higher stability to fabrication imperfections and provide better characteristics of semiconductor microdisk lasers and microwave and optical oscillators [6,21,22]. Besides, high near-field intensity in the region of the notch can possibly be exploited to enhance the sensitivity of photonic biosensors based on the WG-mode resonators [23].

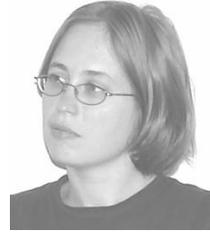

**Svetlana V. Boriskina** (S'96-M'01) was born in Kharkiv, Ukraine in 1973. She received the M.Sc. degree with honours in radio physics and Ph.D. degree in physics and mathematics from Kharkiv National University, Ukraine, in 1995 and 1999, respectively.
From 1997 to 1999 she was a Researcher in the School of Radio Physics at the Kharkiv National University, and in 2000, a Royal Society – NATO Postdoctoral Fellow in the School of Electrical and Electronic Engineering, University of Nottingham, UK. Currently she works there as a Research Fellow. Her research interests are in integral equation methods for electromagnetic wave scattering and eigenvalue problems, with applications to open waveguides, semiconductor microcavity lasers, and optical filters.

**Trevor M. Benson** (M'95-SM'01) was born in Sheffield, England in 1958. He received a First Class honours degree in physics and the Clark Prize in Experimental Physics from The University of Sheffield in 1979 and a PhD in electronic and electrical engineering from the same University in 1982. After spending over six years as a Lecturer at University College Cardiff, Professor Benson joined the University of Nottingham as a Senior Lecturer in Electrical and Electronic Engineering in 1989. He was promoted to the posts of Reader in Photonics in 1994 and Professor of Optoelectronics in 1996. Professor Benson has received the Electronics Letters and JJ Thomson Premiums from the Institute of Electrical Engineers. He is a Fellow of the Institute of Electrical Engineers (IEE) and the Institute of Physics. His present research interests include experimental and numerical studies of electromagnetic fields and waves, with particular emphasis on propagation in optical waveguides and lasers, glass-based photonic circuits and electromagnetic compatibility.

**Phillip Sewell** (S'88-M'91-SM'04) was born in London, England in 1965. He received the B.Sc. Degree in electrical and electronic engineering from the University of Bath with first class honours in 1988 and the Ph.D. degree from the same university in 1991. From 1991 to 1993, he was an S.E.R.C. Postdoctoral Fellow at the University of Ancona, Italy. Since 1993, he has been with the School of Electrical and Electronic Engineering at the University of Nottingham, UK as Lecturer, Reader (from 2001) and Professor of Electromagnetics (from 2004). His research interests involve analytical and numerical modeling of electromagnetic problems, with application to optoelectronics, microwaves and electrical machines.

**Alexander I. Nosich** (M'94-SM'95-F'04) was born in Kharkiv, Ukraine in 1953. He graduated from the School of Radio Physics of the Kharkiv National University in 1975. He received Ph.D. and D.Sc. degrees in radio physics from the same university in 1979 and 1990, respectively. Since 1978, he has been with the Institute of Radio-Physics and Electronics (IRE) of the National Academy of Sciences of Ukraine, in Kharkiv, where he holds a post of Leading Scientist in the Department of Computational Electromagnetics. Since 1992 he held research fellowships and visiting professorships in the EU, Turkey, Japan, and Singapore. His interests include methods of analytical regularization, free-space and open-waveguide scattering, complex mode behavior, radar cross-section analysis, modelling of laser microcavities, and antenna simulation.